\documentclass[superscriptaddress,floatfix,reprint, amsmath,amssymb,
aps,pre]{revtex4-2}
\usepackage[margin=1in]{geometry}
\usepackage{graphicx}
\usepackage[siunitx]{circuitikz}
\usepackage{float}
\usepackage{comment}

\usepackage{soul}

\usepackage{subcaption}

\renewcommand{\hl}[1]{#1}

\begin{document}
		
\title{Experimental measurements of the granular density of modes via impact}
\author{Sydney A. Blue}
\affiliation{Physics Department, Presbyterian College, Clinton, 29325 South Carolina, USA}
\author{Salem C. Wright}
\affiliation{School of Materials Science and Engineering, Georgia Institute of Technology, Atlanta, 30332 Georgia, USA}
\author{Eli T. Owens$^1$}


\begin{abstract}
The jamming transition is an important feature of granular materials, with prior work showing an excess of low frequency modes in the granular analog to the density of states, the granular density of modes. In this work, we present an experimental method for acoustically measuring the granular density of modes using a single impact event to excite vibrational modes in an experimental, three dimensional, granular material.  We test three different granular materials, all of which are composed of spherical beads.  The first two systems are monodisperse collections of either 6~mm or 8~mm diameter beads.  The third system is a bidisperse mixture of the previous two bead sizes. During data collection, the particles are confined to a box; on top of this box, and resting on the granular material, is a light, rigid sheet onto which pressure can be applied to the system.  To excite the material, a steel impactor ball is dropped on top of the system.  The response of the granular material to the impact pulse is recorded by piezoelectric sensors buried throughout the material, and the density of modes is computed from the spectrum of the velocity autocorrelation of these sensors.  Our measurements of the density of modes show more low frequency modes at low pressure, consistent with previous experimental and numerical results, as well as several low frequency peaks in the density of modes that shift with applied pressure. Our method represents an experimentally simple technique for investigating the granular density of modes and may increase the accessibility and number of such measurements.  
\end{abstract}

\maketitle

\section{Introduction}

Jamming is the phenomena in which a granular material becomes rigid and able to support a finite shear stress and pressure~\cite{Liu1998,Cates1998,OHern2003, Majumdar2007, Hecke2010, Behringer2019}.  The jamming transition is important for understanding not only the fundamental behavior of granular materials but can also be applied in diverse industrial settings ranging from universal, soft, robotic grippers~\cite{Brown2010, Shintake2018, Howard2022} to hopper flow~\cite{Rhodes2008,Janda_2009, Kiwing2005} to the manufacture of pharmaceuticals~\cite{Muzzio2002}. Jamming also has application on a wide variety of geophysical processes~\cite{Chang2022, Holtz2011} such as earthquakes~\cite{Sammis2021, Ciamarra2009, Ciamarra2010} and avalanches~\cite{Wang2015}.

The jamming transition exhibits features of a second order phase transition; specifically, diverging length scales are observed as jamming is approached~\cite{Silbert2005,Schwarz_2006,Henkes2005,OHern2003, Ellenbroek2006, Somfai2007}. In particular, simulations of the granular density of states~\cite{Silbert2005, OHern2003, Henkes2010, Wyart2005B}  observe long wavelength modes that diverge as the jamming transition is approached.  These simulations of the density of states, $D(\omega)$, display an excess number of low frequency modes as compared to Debye scaling~\cite{Mizuno2017}, \hl{$\omega^{d-1}$, where $d$} is the dimensionality of the system \hl{and $\omega$ is the frequency}. These simulations identify the diverging length scale by first identifying a characteristic frequency, $\omega^*$, that is either the frequency at which the density begins to fall to zero at low frequency~\cite{Silbert2005}, or the frequency at which $D(\omega)$ falls to some fixed fraction of its maximum value~\cite{Henkes2010} at low frequency. Regardless of how $\omega^*$ is defined, it is observed to have a power-law relationship with proximity to the jamming transition as quantified by excess coordination number $(z-z_c)$ and relative packing fraction $(\phi-\phi_c)$~\cite{Silbert2005, Henkes2010, Somfai2007}. 

In this work, we present a technique for experimentally measuring the granular density of states. We thus begin by briefly summarizing some of the previously used methods for measuring and calculating $D(\omega)$. Simulations calculate the density of states using the dynamical matrix constructed from the particle positions and grain interaction potential. A related method uses a covariance matrix; this method has been used to experimentally measure $D(\omega)$ in a two-dimensional (2D) colloidal systems~\cite{Ghosh2010,Chen2010} by utilizing particle tracking to construct the covariance matrix from the particle displacements induced from thermal motion. However, applying this method to granular experiments requires an external vibration as a substitute for the thermal vibrations naturally present in colloids. Experiments in a 2D granular system~\cite{Brito2010} applied such an external vibration to their system and used particle tracking to construct a covariance matrix; these experiments found the number of long wavelengths mode increased as jamming was approached.  While the previously mentioned matrix methods are successful in finding the vibrational modes of the system, these types of experimental methods are  generally reserved for 2D systems that are  small enough for the particles to be tracked. This presents an impractical option for large 3D systems. Another method employed in Refs.~\cite{Owens2013,Brzinski2018} uses the spectrum of the velocity autocorrelation of the particles to measure the granular density of states (VACF method) and can be applied to large, 3D systems. 

As the VACF method described in Ref.~\cite{Owens2013} will form the basis of our work, we will describe this method in detail. First, the velocity autocorrelation function, $C_v(t)$, is defined as follows
\begin{equation}
C_v(t) = \frac{\sum_{j}\langle v_j(\tau + t)\cdot v_j(\tau)\rangle_\tau}{\sum_{j}\langle v_j(\tau)\cdot v_j(\tau)\rangle_\tau}
\label{eq:C}
\end{equation}
where $v_j(t)$ is the velocity as a function of time of the $j$ oscillator in the system. $v_j(t)$  can be written as 
\begin{equation}
v_j(t)=A_j\omega_j\sin(\omega_j t+\phi_j)
\label{eq:vj}
\end{equation}
where $A_j$ is the amplitude of the oscillation, $\omega_j$ is the angular frequency of the oscillator, and $\phi_j$ is the phase of the oscillator. The sum in Equation~\ref{eq:C} is over all the oscillators in the system. We now briefly summarize the derivation presented by Ref.~\cite{DICKEY1969} in order to clearly identify the assumptions and conditions of this method. The first step of this derivation substitutes $v_j$, as defined in Equation~\ref{eq:vj}, into $C_v(t)$, as defined in Equation~\ref{eq:C}. $C_v(t)$ is then simplified using the following assumptions: the first assumption is that the system is in a steady state such that $A_j$ does not vary with time. The second assumption is that the equipartition theorem is applicable to the system, meaning that each oscillator has an energy, on average, of $kT$, where $k$ is the Boltzmann constant and $T$ is the temperature.  The third assumption is that the phase of the oscillators are random. Using these assumptions, $C_v(t)$, simplifies to 
\begin{equation}
C_v(t)  = \frac{1}{N}\sum_{j} \cos(w_jt)\approx\int_0^\infty D(\omega) \cos(\omega t ) \mathrm{d}\omega 
\label{eq:C2}
\end{equation}
where the sum has been approximated by an integral. From here, it is clear that the density of states is given by the Fourier transform of $C_v(t)$
\begin{equation}
	D(\omega) = \int_{0}^{\infty} C_v(t) e^{-i\omega t}  \mathrm{d}t
	\label{eq:DOM}
\end{equation}
As seen from the above derivation, there are some key assumptions made about the system when using the VACF method to find the density of states. Specifically, this method is valid for a system meeting the following criteria: the equipartition theorem applies, the system is in thermal equilibrium, and the vibrations are isotropic with random phase. This method also assumes access to the velocities of all the particles. 

Given these assumptions, the applicability of this method to granular systems is not guaranteed. However, the VACF method was experimentally applied to a 2D granular packing in Ref.~\cite{Owens2013} with the authors calling the result the ``density of modes'' to distinguish it from the thermal density of states. In these experiments, the granular packing was excited with a steady white noise acoustic signal to mimic thermal motion. However, the system demonstrated non-thermal motion, violating a key assumption of the VACF method. In spite of this, the experiments recovered several important features of the density of state, such as Debye scaling in ordered, crystalline packings. It also saw an excess number of low frequency modes compared to Debye scaling in disordered packings and was able to define an $\omega^*$ that scaled with $z-z_c$ but which scaled only weakly with pressure. These results demonstrate the utility of applying the VACF method to granular systems.   

\begin{figure}
	\begin{subfigure}[b]{0.36\textwidth}
	\includegraphics[trim=75 100 75 200, width=\textwidth]{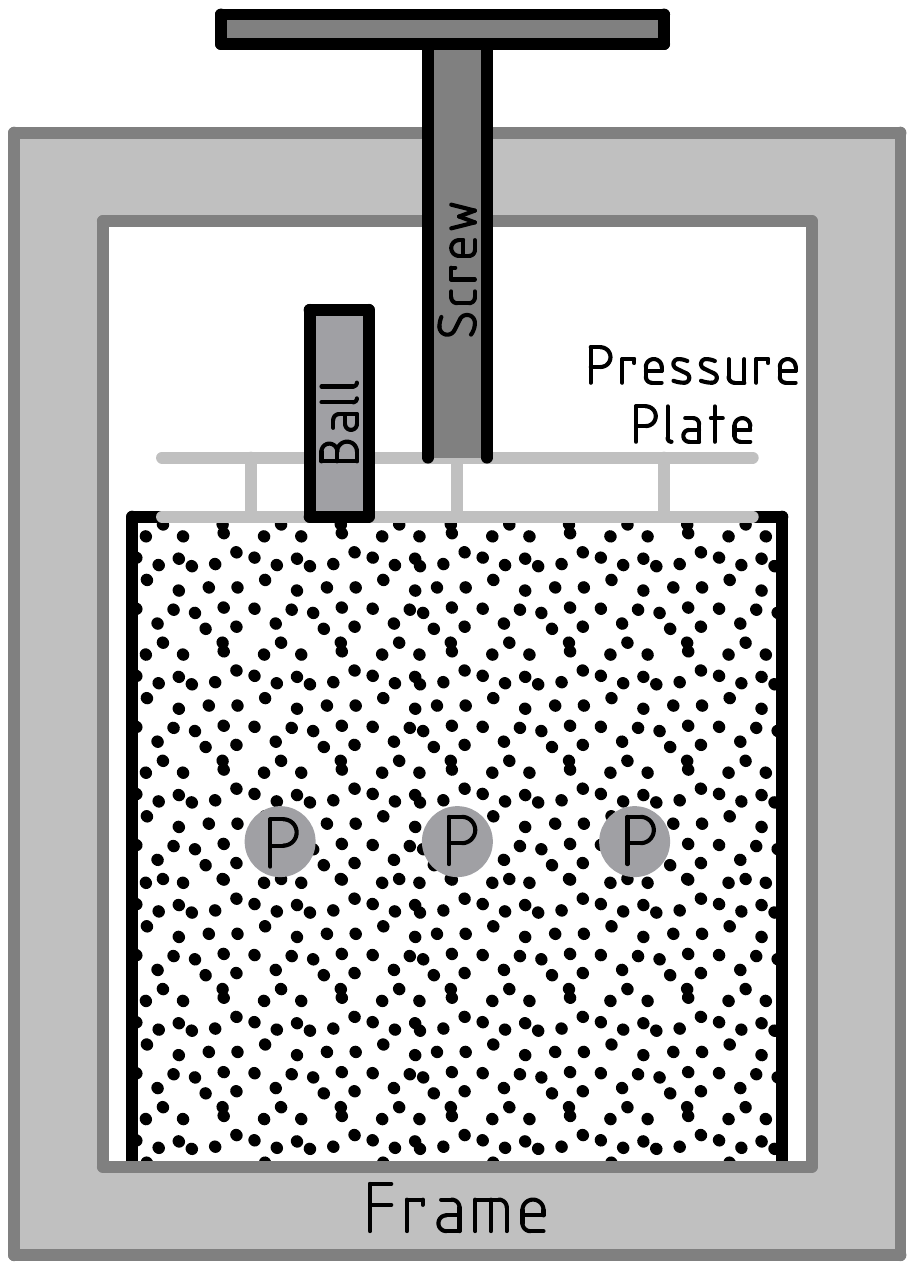}
	\caption{}
	\end{subfigure}
	\begin{subfigure}[b]{0.11\textwidth}
	\includegraphics[angle=90,width=\textwidth]{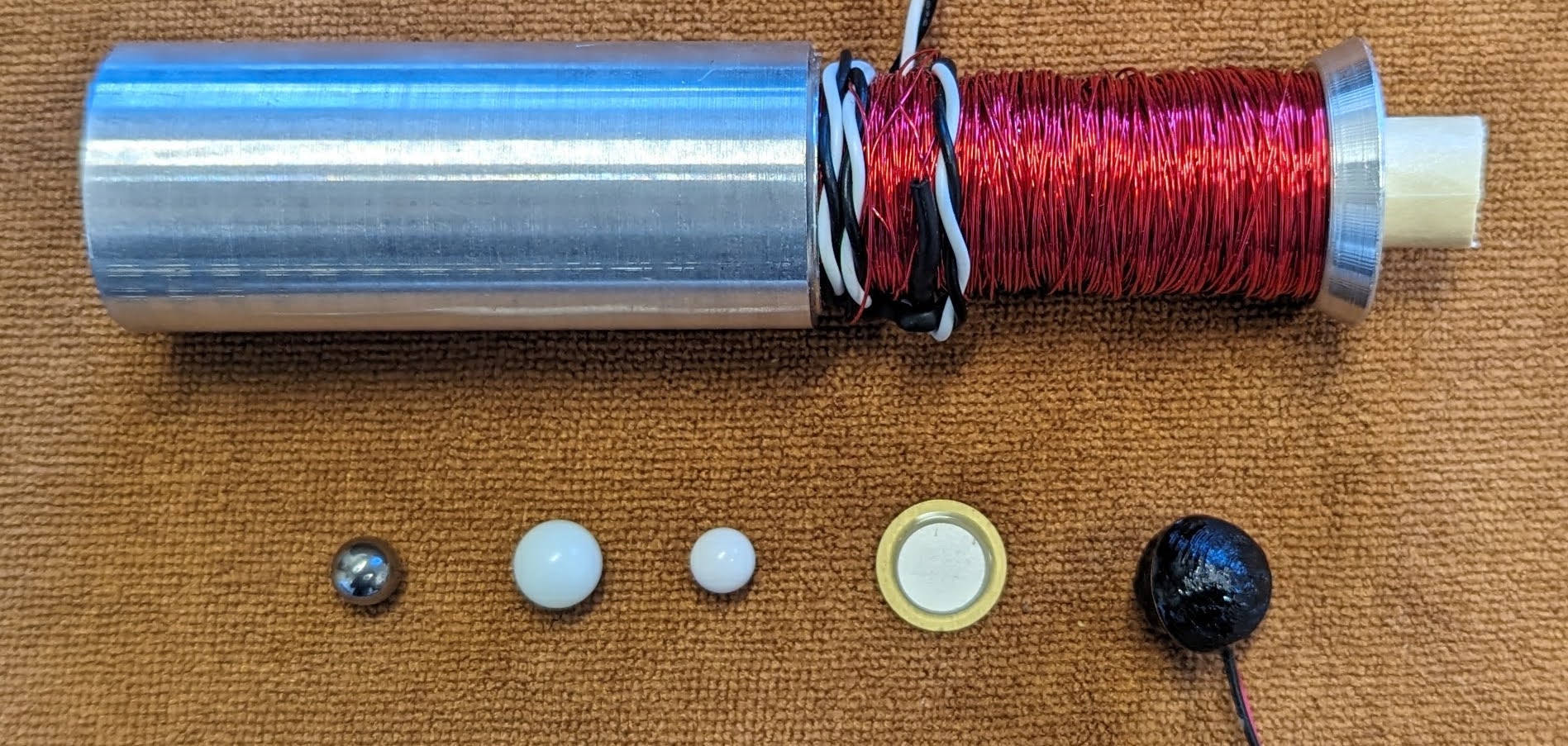}
	~\vspace{3.5em}~
	\caption{}
	\end{subfigure}		
		\caption{(a) A schematic of the experimental setup. The granular material, shown as the dotted fill, is confined inside a box, shown as the black line surrounding the granular material. Inside the granular material, labeled with a ``P'' in the figure are the piezoelectric sensors. On top of the granular material is the pressure plate assembly. The pressure plate assembly houses the force sensors that measure the pressure applied to the granular material via the screw which is itself confined to the frame. Vibrations are excited in the granular material when the ball is released from the solenoid and dropped through a tube onto the bottom plate of the pressure plate assembly. The bottom plate of the pressure plate assembly is in direct contact with the top of the granular material. (b) A photo showing the ball dropping apparatus with the red wires composing the solenoid and the aluminum tube below the solenoid. To the right, and from the top down, are as follows: the piezoelectric sensor encased in the ABS half spheres, the bare piezoelectric sensor, a 6 mm grain, a 8 mm grain, and the steel impactor ball. }	
			\label{fig:setup}
\end{figure}

A key aspect of applying the VACF method to a granular material is to excite the material to create particle velocities that are not normally present in a granular material. In Ref.~\cite{Owens2013}, the granular material was excited with a white noise acoustic signal to mimic thermal motion. This method of exciting the system can be thought of as similar to a damped, driven oscillator where the system is forced to oscillate at the drive frequencies, but the amplitude of these oscillations are attenuated in relation to the natural resonances of the system. That is, the system will attenuate frequencies that are not resonant with the system while minimally attenuating resonant frequencies. This method of exciting the system has an advantage that dissipation in the granular material is counteracted due to the constant white noise drive. However, there are several significant experimental drawbacks related to the complexity of implementing this method. Specifically, a very flat spectrum of white noise must be injected into the system due to the fact that any electrical or mechanical resonances in the driving apparatus will be transmitted to the density of states. Compensating for these resonances requires careful design of electronics and calibration making this a difficult excitation method to apply outside of carefully controlled laboratory conditions. Additionally, since the system is being continuously driven, a relatively large and powerful driving mechanism must be used.  An alternative method of exciting the vibrational modes that used the VACF method is reported in Ref.~\cite{Brzinski2018} where they used the natural phonon emissions of stick slip events of a granular material under shear to compute the density of modes. This method eliminated the above mentioned experimental difficulties in the driving method; however, this method is not applicable to static granular packings due to the lack of shear.

Drawing from the work of Refs.~\cite{Owens2013,Brzinski2018}, we propose a method to excite vibrational modes in a granular material for the purpose of measuring the granular density of modes with the VACF method. Our method excites the vibrational modes using a small impact on top of the granular material. This impact will create a pulse containing a wide spectrum of frequencies. The range of frequencies in a pulse is inversely proportional to the width of the pulse, with an infinitesimal pulse having a flat spectrum. This wide injection of frequencies will allow the granular material to vibrate more strongly at its natural resonances and attenuate signals that are not resonant with the natural modes of the system.

This impact method of excitation is experimentally simple; it does not require complex electronics, large shakers, or the need to compensate for apparatus resonances. In addition to being experimentally simple, an impact has a particular intrinsic advantage. Namely, while the impactor is striking the granular material, the system can also be thought of as a damped, driven oscillator; however, since the impact event is very short, transients will be important to the response. For damped, driven oscillators, the transient terms oscillate at the natural frequency of the system and not the driving frequency, which is exactly what is needed for our experiments. Then, after the impact event, the driving force is no longer present and the system behaves more like a damped oscillator, essentially ringing at the natural frequencies of the material. This allowance of transients is in contrast to previously used continuous excitation methods~\cite{Owens2013} where the system is forced to oscillate at all modes and then attenuate modes that are not resonant with the system. The primary disadvantage of this impact technique is also due to the transient nature of the excitation, in that, it may be difficult to overcome dissipation as the vibrations travel through the material.  

In this paper, we experimentally measure the granular density of modes of a real three-dimensional granular material utilizing the VACF method with an impact event to excite the vibrational modes. Our results recover several previously discussed features of the granular density of states, such as an excess of low frequency modes at low pressure. Additionally, this work adds another experimental system to the relatively small number of systems for which there is data on the granular density of modes.

\section{Experimental Methods}

For our experiment, we investigate a granular material confined within a box closed on five sides and open at the top, see Figure~\ref{fig:setup}. The granular material inside the box has dimensions of $25\times25\times25$~cm$^3$ and is composed of one of the following three possible grain types: i) a monodisperse packing of 6 mm spherical grains, ii) a monodisperse packing of 8 mm spherical grains, iii) a 50-50 bidisperse mixture by volume of 8 mm and 6 mm spherical grains. For both grain sizes, the grains are composed of ABS plastic with a Young's modulus of $E\approx2.6$~GPa~\cite{ABS}, a Poisson's ratio of $\nu\approx0.37$~\cite{WebMat, Zou2016}, and individual grain masses of $m_{6mm}=0.20$~g and $m_{8mm}=0.34$~g.

In order to shift the granular material further from the jamming transition, pressure is applied to the granular material via a pressure plate apparatus placed on top of the granular system. This pressure plate apparatus consists of two $25\times25$~cm$^2$ parallel plates with three force sensors (TE Connectivity Measurement Specialties, FX19 Series sensors) placed in between the plates to allow us to measure the total force applied to the system. Pressure is then applied to the pressure plate apparatus via a screw built into a frame that encircles the grain box, see Figure~\ref{fig:setup}. This apparatus for applying pressure to the granular material is similar to a cider press for pressing apples into juice, where the grains are the ``apples'' in this analogy. For our experiments, the lowest pressure is achieved with no force applied from the screw; then, we apply the following pressures to the system by tightening the screw and compressing the granular material from the top: 0~kPa, 3.6~kPa, 7.1~kPa, 14.2~kPa. Since the sensors are buried in approximately the middle of the granular packing, we add the weight of the grains above the sensors (50 N) for a revised pressure range of 0.8~kPa,  4.4~kPa, 7.9~kPa, 15~kPa. For a more intuitive understanding of these pressure ranges, the pressures can be expressed as fractions of the Young's modulus of the grains as follows: $0.3~\mu E$, $1.7~\mu E$,  $3.0~\mu E$, $5.8~\mu E$, where the unit $\mu E=10^{-6}E$.

\begin{figure}
	{\centering	
	\resizebox{0.5\textwidth}{!}{%
		\begin{tikzpicture}
			\coordinate (P+) at (0,2);
			\coordinate (P-) at (0,-2);
			\coordinate (R+) at (4,1.5);
			\coordinate (R-) at (4,-1.5);
			
			\draw (P-) to[PZ,l_=piezo] (P+); 
			
			\draw (R+) node[op amp] (A1) {$A_1$};
			\draw (P+) to[R,l^=$R_i$\SI{=100}{\ohm}] (A1.-);
			\draw (A1.-) to[short,*-] ($(A1.-) + (0,1.5)$) to[C,l_=$C_f$\SI{=1}{nF} ] ($(A1.-) + (2.4,1.5)$) to[short,-*] ($(A1.-) + (2.4,-.5)$);
			\draw ($(A1.-) + (0,1.5)$) to[short,*-] ($(A1.-) + (0,3)$) to[R,l_=$R_f$\SI{=10}{M\Omega} ] ($(A1.-) + (2.4,3)$) to[short,-*] ($(A1.-) + (2.4,1.5)$);
			
			\draw (R-) node[op amp, yscale=-1] (A2) {};
			\draw (R-) node {$A_2$};
			\draw (P-) to[R,l_=$R_i$\SI{=100}{\ohm}] (A2.-);
			\draw (A2.-) to[short,*-] ($(A2.-) + (0,-1.5)$) to[C,l^=$C_f$\SI{=1}{nF} ] ($(A2.-) + (2.4,-1.5)$) to[short,-*] ($(A2.-) + (2.4,+.5)$);
			\draw ($(A2.-) + (0,-1.5)$) to[short,*-] ($(A2.-) + (0,-3)$) to[R,l^=$R_f$\SI{=10}{M\Omega} ] ($(A2.-) + (2.4,-3)$) to[short,-*] ($(A2.-) + (2.4,-1.5)$);
			
			\draw (A1.+) to (A2.+);
			\draw (2.82,0) to[short,*-] (2.3,0) node[ground]{};
			
			\draw (8,0) node[op amp, yscale=-1] (A3)  {};
			\draw (8,0) node {$A_3$};
			\draw (A3.+) to ($(A3.+)+(-.5,0)$) to ($(A3.+)+(-.5,1)$)  to ($(A3.+)+(-1.6,1)$);
			\draw (A3.-) to ($(A3.-)+(-.5,0)$) to ($(A3.-)+(-.5,-1)$)  to ($(A3.-)+(-1.6,-1)$);
			\draw (A3.out) node[shift={(0,.3)}] {$V_{out}$};
			\draw(A3.out) to[short,-o] ($(A3.out)+(0.1,0)$);
		\end{tikzpicture}
	}}
	
	\caption{A conceptual schematic of the piezoelectric pre-processing circuit (details are omitted). $A_1$ and $A_2$ together with $R_f$ and $C_f$ form the two charge amplifiers and set the low frequency response of the system. The two resistors, $R_i$, together with the piezoelectric's capacitance set the high frequency response. Finally, $A_3$ is the differential amplifier that gives the final output voltage proportional to the charge difference on the piezoelectric.  } 
	
	\label{fig:circuit}

\end{figure}
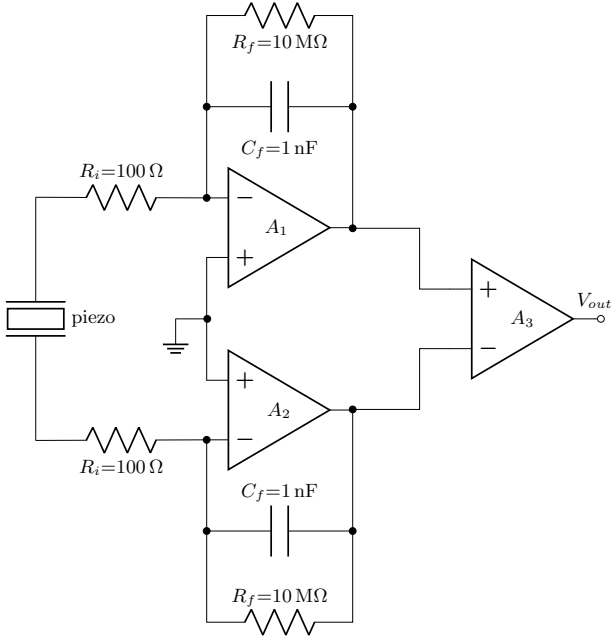

\begin{figure}
	\centering
	\includegraphics[width=0.5\textwidth]{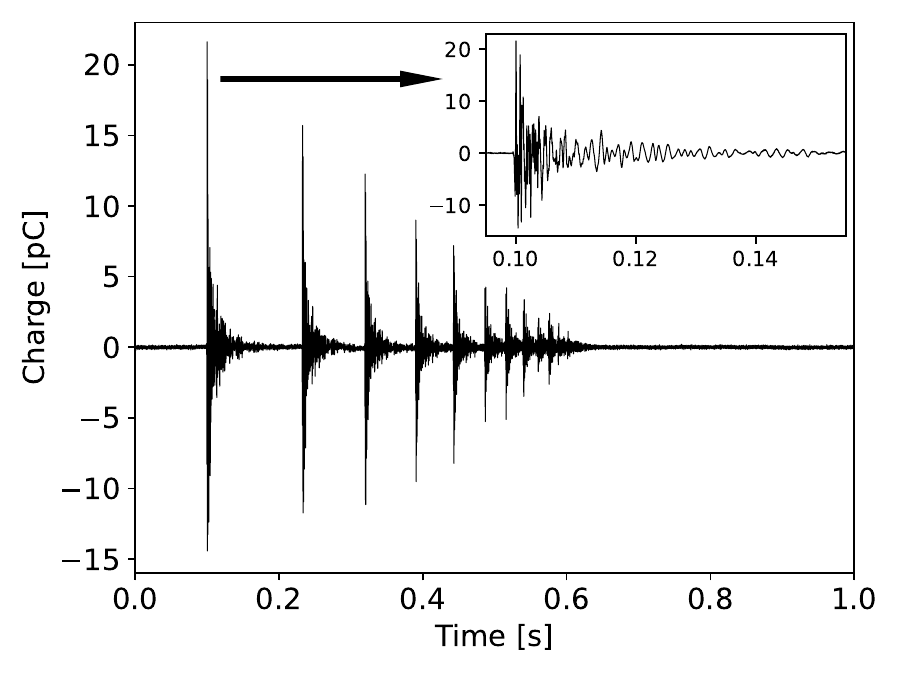}
	\caption{A typical charge versus time signal from one of the piezoelectric sensors. The multiple pulses are due to the ball bouncing on the pressure plate; the time between the first and second bounce is approximately 0.1 seconds. The inset is a zoomed in view of the first pulse. }
		\label{fig:data}
\end{figure}

\subsection{Piezoelectric sensors}

We use five piezoelectric sensors embedded in the granular material to measure the acoustic vibrations of the system caused by the impact of a steel ball being dropped on the granular material. The particular piezoelectric sensors used are Murata 7BB-12-9 buzzers. While these crystals are marketed as buzzers, they work equally well as detectors due to the symmetric nature of the piezoelectric effect: that is, a strain on a piezoelectric crystal will produce charge separation and, inversely, a charge difference across a piezoelectric crystal will produce a strain. The piezoelectric sensors are 12 mm in total diameter with the crystal being 9 mm in diameter. For each sensor, we glue two ABS plastic half spheres on either side in order to turn the sensor into a 12 mm diameter ABS sphere with a piezoelectric sensor running through its equator, see Figure~\ref{fig:setup}. Encasing the sensors in ABS spheres allows the sensor to better blend in with the rest of the material which is also composed of ABS plastic spheres. This helps to ensure that the sensors are vibrating in as close a manner to the bulk granular material as possible.

\begin{figure}
	\centering
	\includegraphics[width=0.5\textwidth]{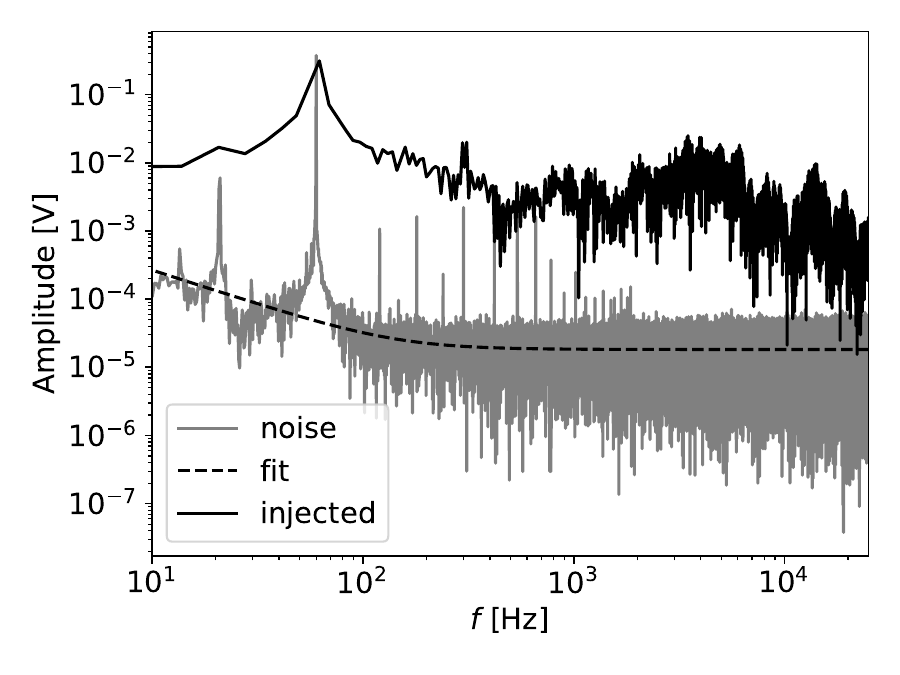}
	\caption{The \hl{unfiltered} spectrum of a piezoelectric affixed to the top of the impact plate showing both the injected frequencies from the impact as well as the noise when there is no impact.  The noise spectrum shows the classic $\frac{1}{f}$ at low frequency followed by broadband noise~\cite{noise}, and we fit the noise spectrum using this form for reference and later use. \hl{The noise spectrum also shows several narrow spikes which are harmonics of the 60~Hz mains. These harmonics of 60~Hz are filtered from the later dataset using a digital notch filter with a bandwidth on the order of 1~Hz.}}
		\label{fig:noise}
\end{figure}

If these sensors are simply placed into the granular material without any additional circuitry, a voltage could be measured in some proportion to acoustic vibrations. However, this signal would be very weak, possibly not even exceeding the noise threshold. It is therefore necessary to connect the piezoelectric sensors to pre-processing circuitry to amplify the signal and reduce the noise level. The circuit that we developed (shown in Figure~\ref{fig:circuit}) to accomplish this task consists of two charge amplifiers~\cite{SLOA033A}, one connected to each side of the crystal. These charge amplifiers convert the charge on each side of the crystal into a voltage. These two voltages are then fed through a differential amplifier for both amplification and common mode noise rejection. This circuit yields high sensitivity and low noise, ideal for measuring small vibrations.  Additionally, the output voltage of the circuit, $V_{out}$ can be converted into the charge, $q$, generated by the piezoelectric with the following relation $q=\frac{V_{out}}{2G C_f}$, where $G$ is the gain of the differential amplifier and $C_f$ is the feedback capacitor. Figure~\ref{fig:data}  shows the output of a typical piezoelectric sensor. 

This circuit also introduces low and high frequency cut offs for our measurements. The low frequency cutoff is set by the feedback resistor and capacitor in the charge amplifiers. For our circuit values, this gives a 16Hz, -3dB frequency. A high frequency cut off is intentionally introduced in order to roll-off high frequency noise that is outside the frequency range of interest, and is set by the input resistance $R_i$ and the capacitance of the piezoelectric crystal (8~nF). Using our circuit values gives a -3dB frequency of 100kHz;  however, in practice, since our sample rate is 50~kHz, we only measure up to 25~kHz due to the Nyquist frequency. 

\subsection{Phonon excitation}

In order to excite the granular material to vibrate at its natural modes, we drop a 1.05~g, 6.35~mm diameter, steel ball, see Figure~\ref{fig:setup}, on top of the system, thereby injecting an acoustic pulse into the system. In order to produce a consistent pulse across our different packings and pressures, the following apparatus was constructed to drop the pulse-producing steel ball on the granular material. The pulse producing apparatus (labeled ``ball'' in Figure~\ref{fig:setup}) consists of a 5.0~cm tall hollow aluminum tube on top of which is a solenoid. Inside and at the top of the tube, but below the solenoid, the steel ball is placed. The steel ball is held in place by the solenoid's magnetic field and can be dropped when the solenoid is turned off. This setup allows for a consistent kinetic energy of the ball  (0.5~mJ) as it strikes the top plate of the system, as well as provides a means of coordinating the time at which the ball is dropped with the data acquisition of the embedded piezoelectric sensors. Additionally, by striking the rigid plate of the pressure plate apparatus, the energy of the pulse is more evenly distributed across the granular material rather than being concentrated in a particular region of the system. 

In order to excite a broad spectrum of frequencies with this impact technique, the impact event should be as short as possible. To design a system that facilities a short impact time, we imagine the impactor colliding with the impactor plate as a mass colliding with a spring, then the period of that spring's oscillation would correspond to the duration of the collision. Using this model for the impact time, a short impact can be accomplished with a light, but hard and stiff impactor. If higher energy is needed from the impactor, it is best accomplished by increasing the speed of the impactor rather than its mass so that the impact time remains small. Finally, so long as the energy of the impactor does not cause the grains to permanently change their position, the energy of the impactor should not influence the $D(\omega)$ of an individual packing, provided there is enough energy to excite the same number of modes with different energies. However, if data from multiple packing are averaged together, as is done in this study, the energy of the impactor should be kept very consistent as packings with higher energy impacts will be weighted more heavily in Equation~\ref{eq:C}.  

Keeping all of these design principles in mind, we chose our impactor to be made of a hardened steel ball bearing with a mass equal to only about five of the 6~mm grains or three of the 8~mm grains. We chose the drop height, and thus the speed of the impactor, by dropping the impactor from the highest height that did not cause the piezoelectric sensor voltages to be clipped. 

\hl{We measure the range of frequencies produced with our impact setup as well as the electrical noise of our circuit by attaching a piezoelectric sensor to the top of the impactor plate and dropping the impactor onto the plate; the results of this test are shown in Figure{~\ref{fig:noise}}. Since the piezoelectric sensor is in close proximity to the impactor, it has a larger amplitude than the piezoelectrics in the granular material, and we  therefore omit the first five bounces of the impactor as the signal from these bounces is clipped by our circuit. Omitting these first bounces simply means that the amplitude of the injected frequencies is greater than what is shown in Figure~{\ref{fig:noise}} and the curve in Figure{~\ref{fig:noise}} represents a floor on the amplitude of the injected frequencies. Additionally, the impactor plate is resting on the granular material and the structure seen in the injected signal is likely due to the granular material. Finally, we see that the injected frequencies are above the noise threshold for the frequencies we investigate, demonstrating that our setup is exciting a broad band of frequencies through the measurement limit of 25~kHz.}

\begin{figure}
	\includegraphics[width=0.5\textwidth]{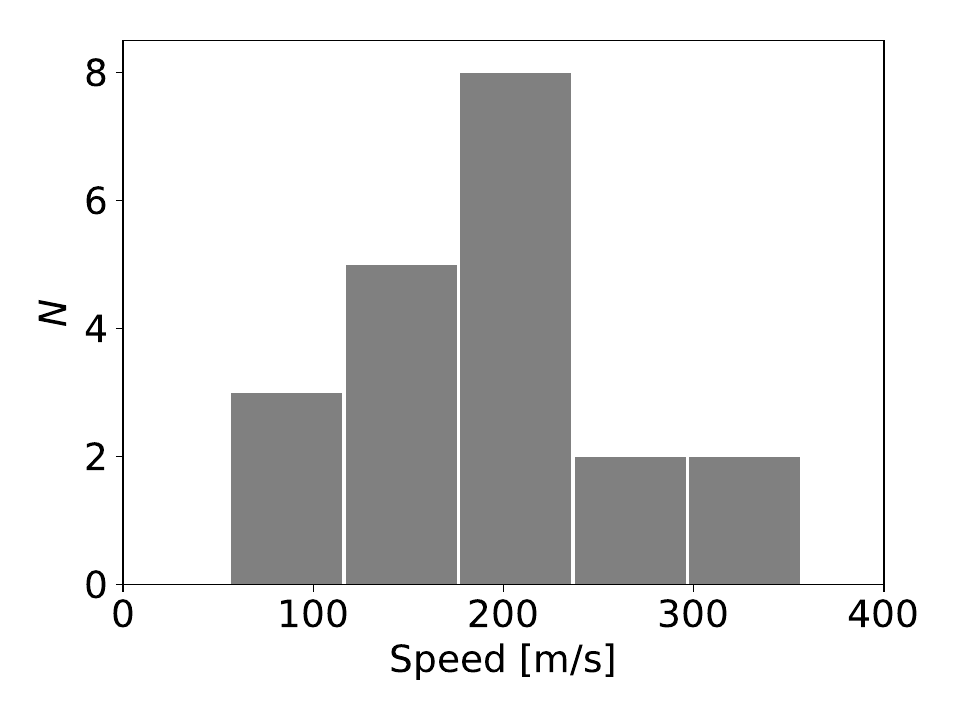}
	\caption{A histogram of the ``time of flight'' sound speeds for the 6~mm monodisperse packing. Each of the 20 speed measurements was conducted on a unique packing and with no confining pressure. The average sound speed was found to be $193\pm16$~m/s (standard error).}
	\label{fig:speeds}
\end{figure}  

\subsection{Data collection protocol}

The data collection procedure is described as follows: First, all the grains are removed from the box, and mixed in a separate container before being poured back into the experimental apparatus. Removing and replacing the grains ensures that we put the granular material into a new, unique state each time we collect data. Then, once the granular material has been placed back into the experimental apparatus, we place the pressure plate on top of the grains, but do not apply any pressure with the screw. At this point, we drop the steel ball onto the system exciting the vibrational modes. At the same time, the solenoid is de-energized and the steel ball is released, the DAQ card is triggered and begins to record the signal from the piezoelectric sensors for one second. Figure~\ref{fig:data} shows a typical output of one of the piezoelectric sensors, with the multiple pulses corresponding to the steel ball bouncing on the pressure plate. 

After this, pressure is applied via the screw to the pressure plate, and the ball is dropped again and the response of the piezoelectrics is again recorded. This procedure is repeated for a total of four pressure states (listed above). Once all the pressure states have been cycled through, the grains are removed, mixed, added back into the experimental apparatus, and the whole procedure is repeated. We measure the density of modes as the pressure on the system is cycled from low to high, since we find these results are not qualitatively different from cycling the pressure on the system from high to low. In total, we perform this procedure on five unique packings for each of our three grain types (15 unique packings in total). 

\begin{figure*}
	
	\begin{subfigure}[b]{0.32\textwidth}
		\includegraphics[width=\textwidth]{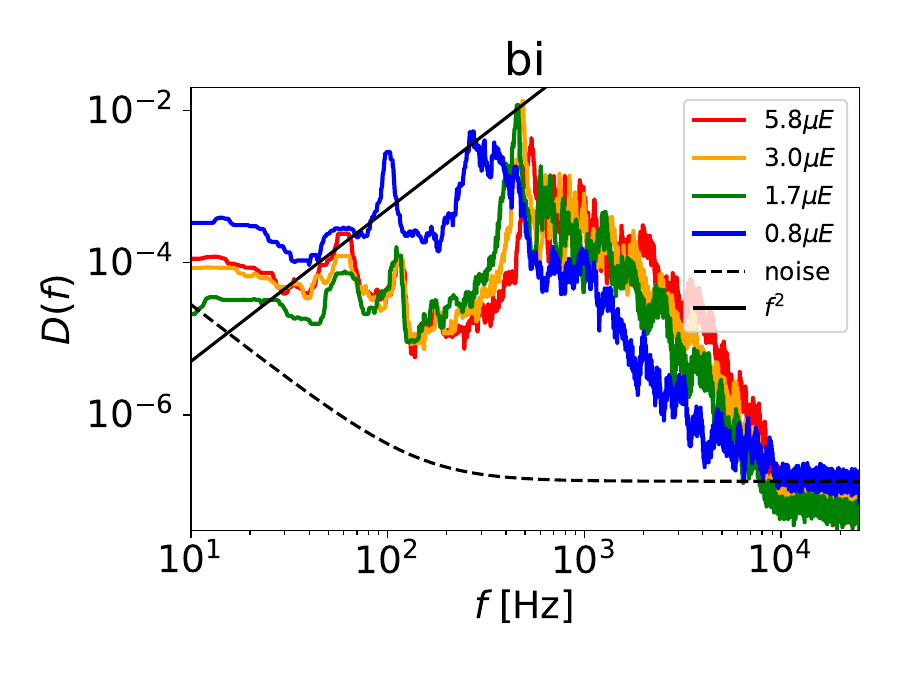}
		\caption{}
	\end{subfigure}
	\hfill
	\begin{subfigure}[b]{0.32\textwidth}
		\includegraphics[width=\textwidth]{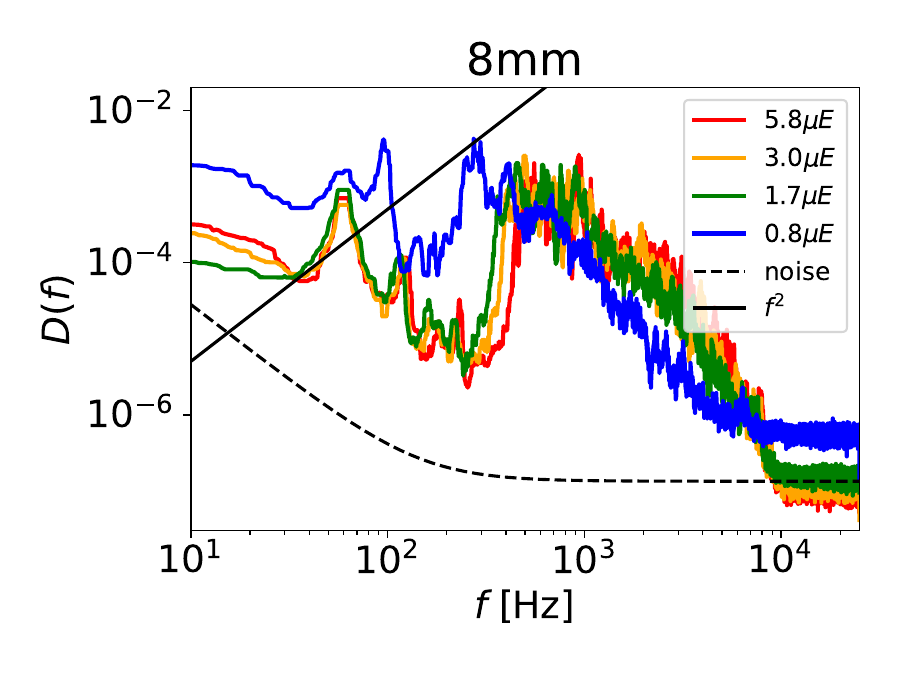}
		\caption{}
	\end{subfigure}
	\hfill
	\begin{subfigure}[b]{0.32\textwidth}
		\includegraphics[width=\textwidth]{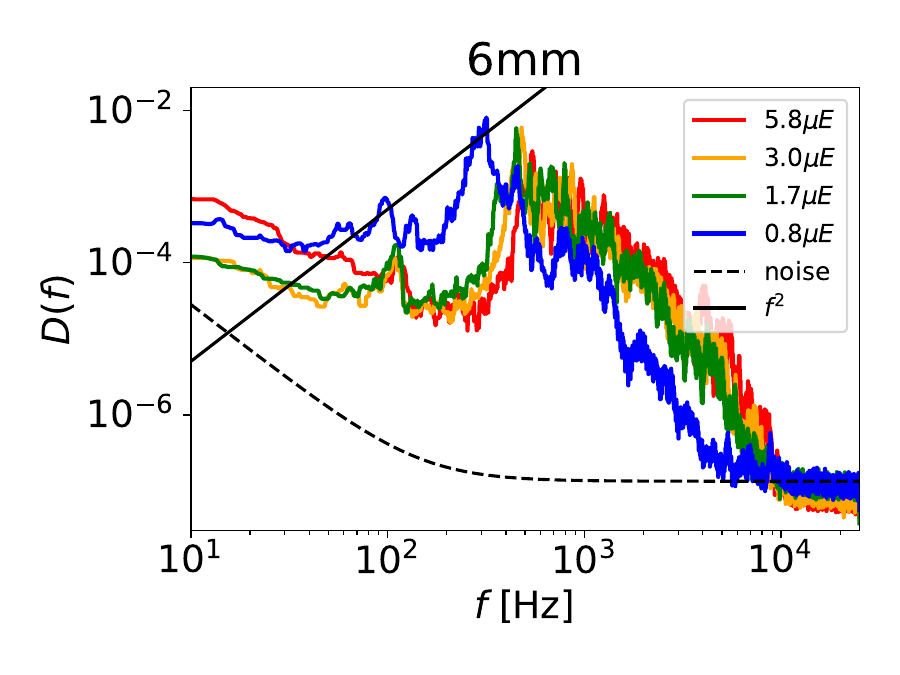}
		\caption{}
	\end{subfigure}
	
	\caption{The density of modes for (a) the bidisperse system, (b) the 8 mm monodisperse system, and (c) the 6 mm monodisperse system. The different colors correspond to the different pressures that the system is subjected to. The dashed line is the power spectrum of the noise fit from Figure~\ref{fig:noise}. The vertical units of the noise power spectrum are arbitrary in order for the noise spectrum to roughly, vertically align with the high frequency noise in the density of modes. \hl{The solid black line is $\propto f^2$ for comparison to Debye scaling.}}
	\label{fig:DOM}
	
\end{figure*}

\subsection{Sound speeds}

In order to be able to relate specific frequencies to relevant length scales, we need to know the speed of sound in our granular material. This was accomplished for the 6~mm packing by measuring the time difference between the first arrival of sound (time of flight) at two piezoelectric sensors separated from each other by 10 cm. For these sound speed measurements, one piezoelectric sensor is placed near the top of the system and the other is placed 10 cm below this. A pulse of sound is injected at the top of the system and the time at which the signal first arrives at each sensor is found. From this time of flight information, a sound speed for the material is found. The grains are then agitated to put the material in a new state and the sound speed measurement is repeated for a total of 20 unique grain configurations. A histogram of these speeds is shown in Figure~\ref{fig:speeds}. 

\begin{figure*}
	
	\begin{subfigure}[b]{0.32\textwidth}
		\includegraphics[width=\textwidth]{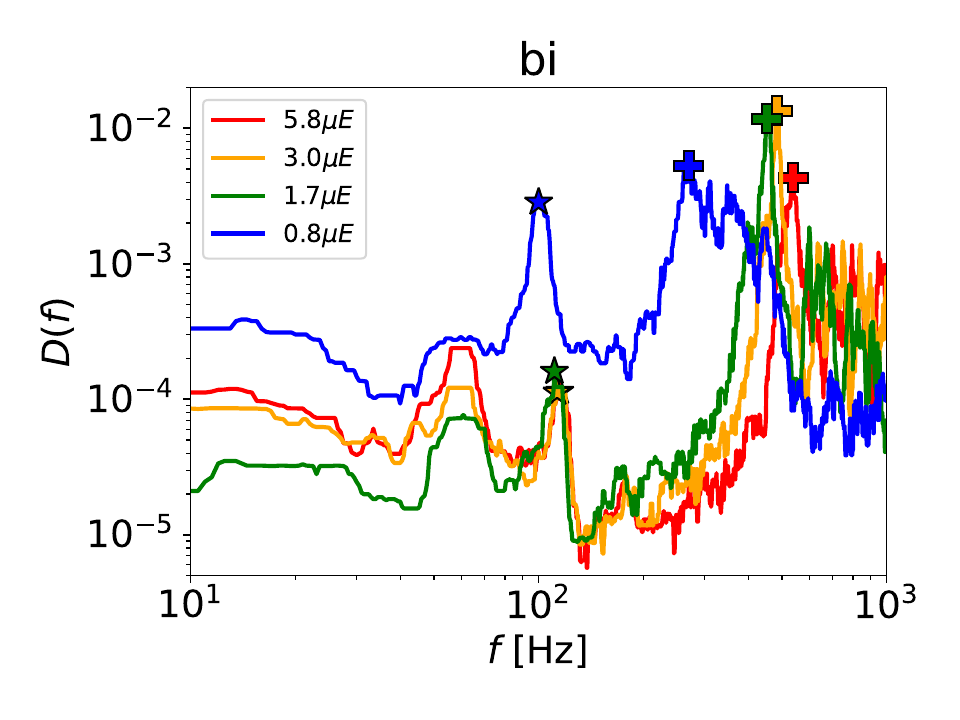}
		\caption{}
		\label{fig:zoom}
	\end{subfigure}
	\hfill
	\begin{subfigure}[b]{0.32\textwidth}
		\includegraphics[width=\textwidth]{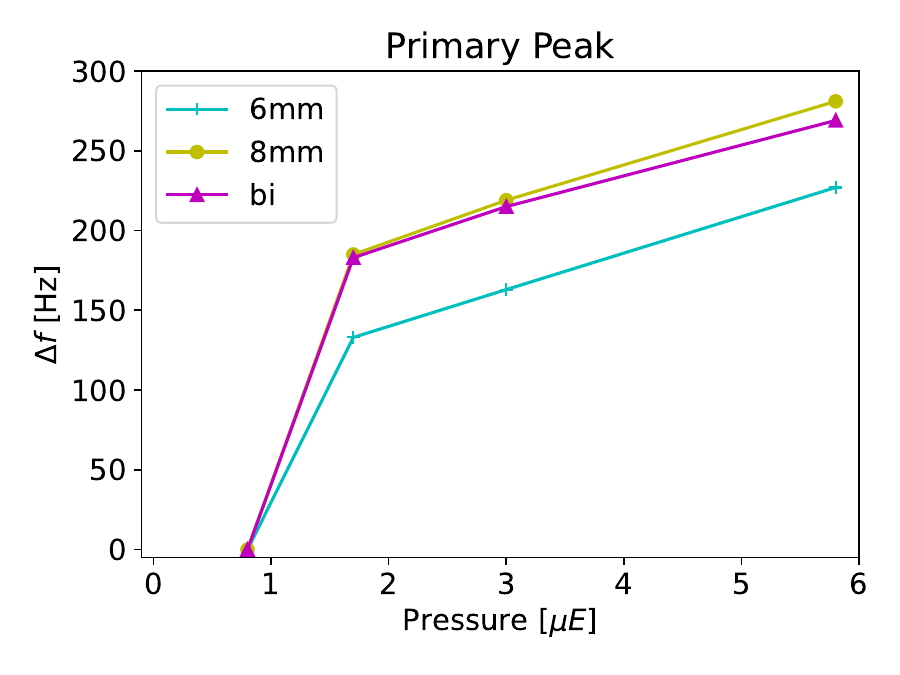}
		\caption{}
		\label{fig:DOM_P1}
	\end{subfigure}
	\hfill
	\begin{subfigure}[b]{0.32\textwidth}
		\includegraphics[width=\textwidth]{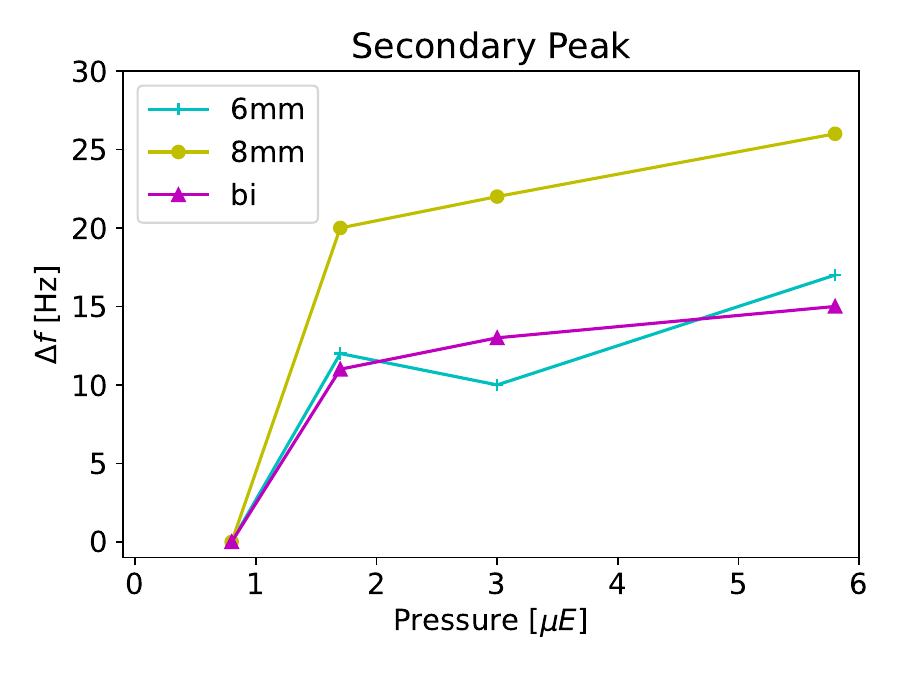}
		\caption{}
		\label{fig:DOM_P2}
	\end{subfigure}
	
	\caption{(a) The DOM for the bidisperse packing zoomed in on the frequencies between 10~Hz and 1~kHz. The location of the primary peaks are labeled with a plus sign and the secondary peak with a star symbol. (b) The $\Delta f$ of the primary peak as a function of pressure. (c) The $\Delta f$ of the secondary peak as a function of pressure.}	
\end{figure*}

We find an average speed of $193\pm16$~m/s (standard error). This speed of sound is significantly slower than the bulk sound speed, $c=\sqrt{\frac{E}{\rho}}$, of 1,200~m/s. This discrepancy with the bulk speed is consistent with previous experiments~\cite{Liu1992,Liu1993,Liu1994} that found the granular speed of sound for a granular material composed of unconfined glass beads to be 280~m/s compared to a bulk speed of 4,000~m/s. Additional experiments on a 2D granular material~\cite{Owens2011} found a wide range of sound speeds within a given packing due to the fact that the speed of sound varies with the strength of the force chain along which the sound propagates.

Additionally, simulations of the sound speed from the time of flight~\cite{Somfai2005} find a pressure dependence of $c\propto P^{1/6}$. In their simulations, they report pressure in units of $E^*=\frac{E}{1-\nu^2}$ and speeds in units of $c^*=\sqrt{\frac{E^*}{\rho}}$, which for our 6~mm packing are $E^*\approx3$~GPa and $c^*\approx1700$~m/s. At the pressure at which our sound speeds were measured ($0.7\mu E^*$), the simulation finds a sound speed of $0.12c^*=200$~m/s in good agreement with our average result of 193~m/s.

\section{Results and Analysis}

Using the piezoelectric charge as a function of time data (see Figure~\ref{fig:data} for an example), we can calculate the granular density of modes via the spectrum of the velocity autocorreclation function. As a first step towards calculating the density of modes, we claim that the the average velocities of the grains around a sensor particle is proportional to the charge on the piezoelectric crystal, that is $q \propto v$ where $q$ is the charge on the crystal and $v$ is the average velocity of the grains around the sensor. Our reasoning for this conclusion is as follows. 

First, we recognize that the charge generated on the piezoelectric is proportional to the strain on the crystal, that is $q \propto  x$, where $x$ is the change in thickness of the crystal. From this, we can see from both an electrical and mechanical perspective that the energy imparted to the crystal from the granular material is proportional to $q^2$. The electrical perspective treats the piezoelectric crystal as a capacitor, and the energy of any capacitor is proportional to $q^2$. The mechanical perspective treats the crystal as a spring with the energy of a spring proportional to $x^2$ which for a piezoelectric is also proportional to $q^2$. 

Second, since the piezoelectric sensors are only in contact with the granular material, all of the energy imparted to the piezoelectric must come from the kinetic energy of its neighboring grains. Since the kinetic energy of the grains is proportional to $v^2$ and the energy of the piezoelectric crystal is proportional to $q^2$, it follows that $v \propto q$ where the constant of proportionality is some combination of material properties of the grains and the crystal, for example Young's modulus, geometry, piezoelectric charge coefficient, etc. The fact that $q$ does not have the precise conversion factor to exactly find $v$ in proper units is not necessary for computing the DOM since Equation~\ref{eq:C} simply needs to know the relative amplitudes of $v$ at each frequency, and then the normalization of the DOM will happen automatically as a consequence of Equation~\ref{eq:C}.

\begin{figure}
	\includegraphics[width=0.5\textwidth]{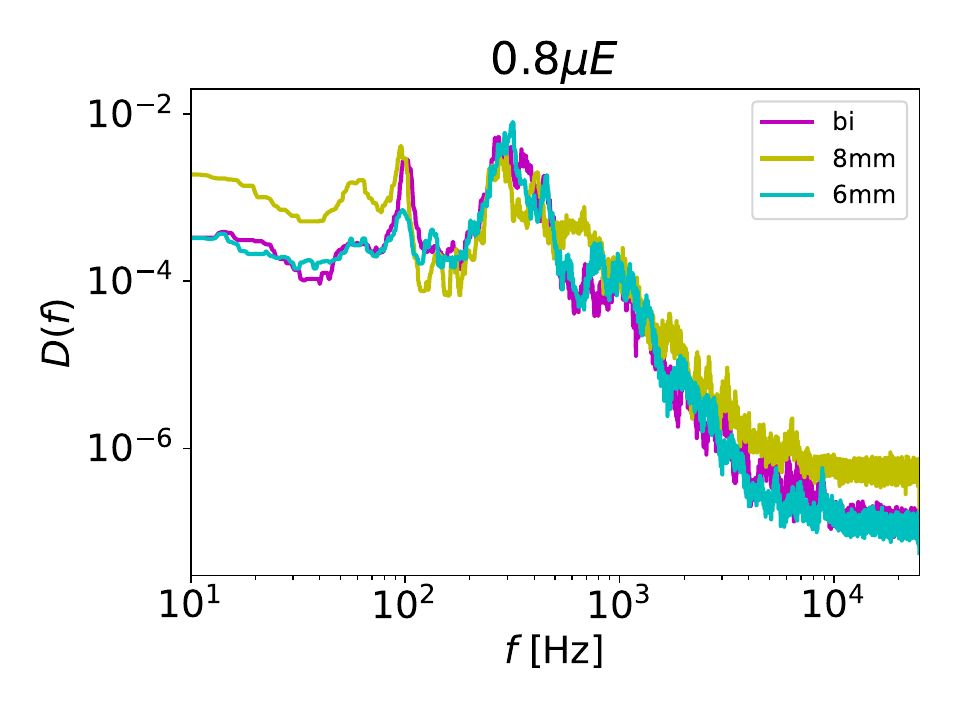}
	\caption{The density of modes for the lowest system pressure for all three grain types. }
	\label{fig:3grains}
\end{figure}  

Now that the velocity of the particles around the sensors can be deduced, the rest of the calculation follows straightforwardly. For each particle type, i.e., 6~mm monodisperse, 8~mm monodisperse, or bidisperse, and at each pressure the velocity autocorrelation function, $C_v(t)$, is found by finding the average charge autocorrelation function for all the sensors. Then, as described in Equation~\ref{eq:DOM}, the FFT of $C_v$ gives the density of modes. Now that we have obtained the density of modes, a final smoothing step is taken by applying a median filter with a 15~Hz window to the density of modes. Compared to the 25~kHz data range, a 15~Hz filter window does not substantially change the density of modes other than a slight flattening of some of the peaks, but it does smooth some of the noise fluctuations seen in the data. The results of these calculations are shown Figure~\ref{fig:DOM}. These results are reported in the frequency range of 10~Hz to 25~kHz. 10~Hz is the lowest frequency that we report due to the fact that the impactor ball has a bouncing period of approximately 10~Hz and the low frequency 3dB point of our piezoelectric circuitry is 16~Hz. The high frequency cutoff of 25~kHz is set by the Nyquist frequency. As a check that our results are above the noise threshold of the electronics, we also add the power spectrum of the noise found in Figure~\ref{fig:noise} to the results in Figure~\ref{fig:DOM}. We add the power spectrum due to Wiener-Khinchin’s theorem~\cite{Champeney1987} that equates the power spectrum with the Fourier transform of the autocorrelation function.  

Some features that are immediately apparent from the density of modes in Figure~\ref{fig:DOM} are the \hl{deviation of the low frequency modes from Debye scaling} as well as a drop off in the number of modes around 10~kHz. We also see several peaks in the approximately 10--1000~Hz range; for a better visual example, Figure~\ref{fig:zoom} is a zoomed in graph of the density of modes in this frequency range. Looking at Figure~\ref{fig:zoom}, there are two peaks of particular interest that shift with system pressure. The first peak is the overall maximum peak in the system, call this the primary peak, the second peak is around 100~Hz, call this the secondary peak, which also shifts with system pressure. The change in the frequency, $\Delta f$, at which each of these peaks occurs relative to the low pressure peak is found and plotted as a function of system pressure in Figs.~\ref{fig:DOM_P1} and \ref{fig:DOM_P2}, and it is seen that the peaks move to higher frequency with increasing system pressure. 

Finally, we also plot the density of modes at the lowest pressure for all three grain types on the same plot in Figure~\ref{fig:3grains} to compare the effects of grain size and compare monodisperse to bidisperse packings.

\section{Discussion}

From Figure~\ref{fig:3grains}, we see that while there are some differences in the density of modes between the three systems, it is not a dramatic difference, with general agreement across the main features of the density of modes. This is not particularly surprising given that the grains are all made of the same material and of the same shape (spherical). Previous experimental measurements in 2D~\cite{Owens2013} did see a difference between monodisperse packings and bidisperse packing due to the crystallization of the 2D packing. However, in our 3D packings, a bidisperse mixture is not required to break up crystallization as it is in a 2D packing~\cite{Perera1999,Robin1999}. Given these observations, we will focus the remainder of our discussion on the 6~mm monodisperse packing; this discussion should also broadly apply to the 8~mm and bi-disperse packings. 

In order to interpret our results, it is helpful to compare with prior simulations of the density of modes~\cite{Silbert2005, OHern2003, Henkes2010, Wyart2005B}. We begin our comparison to simulations by converting the simulation units used in Ref.~\cite{OHern2003} into our experimental units; in these simulations, one unit of frequency defined is $\frac{1}{2\pi}\sqrt{\frac{V_0}{ma^2}}$ where $a$ is the particle diameter, $m$ is the particle mass, and $V_0$ comes from the particle interaction potential, which for spheres gives a force law $F=\frac{V_0}{a^{5/2}}\delta^{3/2}$, where $\delta$ is the particle overlap.  This simulation force law can be equated to the Hertzian contact force law~\cite{johnson1985} $F=\frac{\sqrt{a}}{3(1-\nu^2)}E\delta^{3/2}$. We can now write the simulation unit of frequency in terms of our particle properties as $\sqrt{\frac{aE}{12\pi^{2}m(1-\nu^2)}}$, which for the 6~mm grains is equal to 28~kHz.

Figure~\ref{fig:DOM} shows the density of modes disappearing into the noise around 10~kHz or 0.36 in simulation units, which is a much lower frequency than the more typical simulation roll off frequency of about $2.5$ seen in Ref.~\cite{OHern2003}. One might at first suspect that our experiment is simply not exciting modes above 10~kHz; however, Figure~\ref{fig:noise} shows that we are in fact exciting vibrational modes above 10~kHz. Instead, the system does not ``ring'' as well at these modes and/or these high frequency modes are more strongly dissipated than their low frequency counterparts. Frequency dependent dissipation would be consistent with the viscoelastic nature of ABS plastic~\cite{Lakes_2009, vattathurvalappil2023} and could represent a deviation from the simulated systems that typically model the grains as elastic. Additionally, the viscoelastic damping of the high frequencies may also account for the more gradual falling off of the high frequency side of the density of modes compared to a very steep drop off seen in simulations. While our short impact time combined with a viscoelastic system may limit the highest frequencies accessible for study, much of the information related to jamming occurs in the lower frequency parts of the density of states where we have good resolution down to approximately 0.0036 in simulation units.   

\hl{The primary peak, shown in Figs.{~\ref{fig:zoom}} and{~\ref{fig:DOM_P1}} increases with increasing pressure.} To investigate this further, these frequencies should be correlated to a length scale via the speed of sound. However, this is not a straightforward endeavour as the speed of sound does not have a consistent value as shown in Figure~\ref{fig:speeds}. If the average value of the speed of sound is used, then the primary peak at the lowest pressure has a wavelength of 60~cm which is very close to twice the system size and may correspond to a standing wave across the system. An interesting method to validate this hypothesis would be to create a system that had different dimensions for the length, width, and height of the box and see if multiple peaks in the density of modes emerge. Additionally, the viscoelastic nature of the ABS plastic may also serve to accentuate the primary peak, since without the frequency dependent damping, the density of states may look more flattened at higher frequencies instead of the gradual roll off discussed above. Nonetheless, this primary peak must represent a dominant mode within the system even if its prominence is raised due to viscoelasticity as compared to simulations.

The secondary peak in Figure~\ref{fig:DOM_P2} is also seen to increase with system pressure. Using the average sound speed, the wavelength of the secondary peak in the lowest pressure system is 200~cm, which is longer than the system size. However, these modes may not actually correspond to wavelengths longer than twice the system size as the speed of sound varies considerably. The slowest speed of sound that we measured in Figure~\ref{fig:speeds} is 56~m/s, and if this sound speed is used, the wavelength would be 58~cm which is again close to twice the system size. Within a given packing the sound speed is going to vary depending on the strength of the force chain on which the sound propagates~\cite{Owens2011}. Given that force distribution in a granular material is dominated by a majority of low force particles~\cite{Majmudar2005}, it is possible that this secondary peak corresponds to these low speed modes. 

The observation of these peaks in the density of modes is a departure from what is seen in simulations. However, the fact that these peaks do change with system pressure gives us confidence that they are real features of the granular material and not spurious electrical noise or artifacts. This can be compared to another peak of note in the density of modes seen at 60~Hz. This peak is slightly flattened due to the median filter and the logarithmic scaling of the figure but is well centered at 60~Hz regardless of system pressure. This peak is almost certainly an electrical artifact of the 60~Hz mains in the building's electrical wiring.

\section{Conclusion}

\hl{This work makes two meaningful contributions to the field of granular physics.} The first and main contribution is the suggestion and experimental exploration of a technique for exciting the vibrational modes of a granular material via impact in order to measure the granular density of modes. The second contribution of this work is to measure the density of modes in an as yet unmeasured, experimental, three dimensional granular system, ABS plastic spheres; thereby providing an additional experimental data point on the nature of the granular density of modes in real materials.

Our measurements of the density of modes displays an excess of low frequency modes at low pressure consistent with previous work. We also identify several interesting peaks that shift with pressure. Future work might vary the dimensions of the granular system in order to see if these peaks shift or split into multiple peaks as the symmetry of the system size is broken. Additionally, we see the density of modes rolls off on the high frequency side at a slower rate and lower frequency than what is seen in simulation. This deviation from simulation is likely due to the viscoelastic nature of the particles and represents an additional complexity of real materials that is not currently captured in the simulations. In conclusion, our approach of exciting the grains with a simple impact increases experimental accessibility and will allow for measurements of the granular density of modes to become more commonplace.

\begin{acknowledgments}
We would like to thank the NASA South Carolina Space Grant Consortium and the South Carolina Independent Colleges and Universities for financial support of this work. We would also like to thank Karen Daniels for helpful comments and discussion on this work as well as Thomas Owens for helpful comments and discussions on the experimental apparatus.
\end{acknowledgments}


%

\end{document}